\documentclass[conference,letterpaper]{IEEEtran}
\usepackage{epsfig}
\usepackage{times}
\usepackage{float}
\usepackage{afterpage}
\usepackage{mathrsfs}
\usepackage{amstext}
\usepackage{amssymb,bm}
\usepackage{latexsym}
\usepackage{color}
\usepackage{graphicx}
\usepackage[cmex10]{amsmath}
\usepackage{amsthm}
\usepackage{cite}
\usepackage{tikz}
\usepackage{tikzscale}
\usepackage{graphicx}
\usepackage[center]{caption}
\usepackage{pstricks}
\usepackage{subfigure}
\usepackage{booktabs}
\usepackage{multicol}
\usepackage{lipsum}
\usepackage{dblfloatfix}
\usepackage{algorithm} 
\usepackage[noend]{algpseudocode} 
\usepackage[normalem]{ulem}
\usepackage{enumerate}
\usepackage{bbm}
\usepackage{dsfont}
\usepackage{tikz}

\newcommand{\E}{\mathbb{E}}

\newcommand{\mbb}{\mathbb}
\newcommand{\mcal}{\mathcal}
\newcommand{\msf}{\mathsf}
\newcommand{\C}{\mathsf{C}}

\newtheorem{thm}{Theorem}

\newtheorem{rem}{Remark}

\newtheorem*{defin*}{Definition}

\algnewcommand\algorithmicforeach{\textbf{for each}}

\algdef{S}[FOR]{ForEach}[1]{\algorithmicforeach\ #1\ \algorithmicdo}

\input{widebar_hack.tex}

\begin{document}

\title{Gaussian 1-2-1 Networks with Imperfect Beamforming} 
\author{
\IEEEauthorblockN{Yahya H. Ezzeldin$^\dagger$, Martina Cardone$^{\star}$, Christina Fragouli$^{\dagger}$, Giuseppe Caire$^*$}
$^{\dagger}$ UCLA, Los Angeles, CA 90095, USA,
Email: \{yahya.ezzeldin, christina.fragouli\}@ucla.edu\\
$^{\star}$ University of Minnesota, Minneapolis, MN 55404, USA,
Email: cardo089@umn.edu\\
$^*$ Technische  Universit\"{a}t  Berlin,
Berlin, Germany, 
Email: caire@tu-berlin.de
}
\IEEEoverridecommandlockouts
\maketitle

\begin{abstract}
In this work, we study bounds on the capacity of full-duplex Gaussian 1-2-1 networks with {\em imperfect} beamforming. In particular, different from the {\em ideal} 1-2-1 network model introduced in~\cite{EzzeldinISIT2018}, in this model beamforming patterns result in side-lobe leakage that cannot be perfectly suppressed. The 1-2-1 network model captures the directivity of mmWave network communications, where nodes communicate by pointing main-lobe ``beams'' at each other. We characterize the gap between the approximate capacities of the imperfect and ideal 1-2-1 models for the same channel coefficients and transmit power. We show that, under some conditions, this gap only depends on the number of nodes. Moreover, we evaluate the achievable rate of schemes that treat the resulting side-lobe leakage as noise, and show that they offer suitable solutions for implementation.
\end{abstract}
\section{Introduction}
Millimeter Wave (mmWave) communications are increasingly penetrating a number of network applications, that range from private networks, mmWave mesh network backhauls, to V2X services and military hotspot applications~\cite{Choi-Heath16,Mueck16,Hur-13,Woo-16}. Accordingly, understanding capacity bounds and developing efficient operational schemes for mmWave networks have become increasingly important.
In this paper, we expand our recent investigation of networks with mmWave nodes equipped with perfect directional beams in~\cite{EzzeldinISIT2018,EzzeldinISIT2019}, and study how the capacity is affected when the beamforming is not ideal, specifically when side-lobe leakage cannot be~suppressed.

In~\cite{EzzeldinISIT2018}, we introduced the Gaussian 1-2-1 network, a model that abstracts the directivity aspect of mmWave communications by assuming that perfect main-lobe beamforming beams, with no side-lobe leakage, are available at the nodes.
We used this {\em ideal} model to study the Shannon capacity in arbitrary network topologies that comprise full-duplex mmWave nodes, i.e., nodes that can receive and transmit simultaneously using two highly directive perfect beams. In particular, in~\cite{EzzeldinISIT2018} we proved that the Shannon capacity of a Gaussian 1-2-1 network with full-duplex nodes can be approximated to within a universal constant gap\footnote{Constant gap refers to a quantity that is independent of the channel coefficients and operating SNR, and solely depends on the number of nodes.}. We use {\em approximate capacity} to coin such an approximation in the remainder of the paper.



The focus of this paper is on Gaussian 1-2-1 full-duplex networks where instead of perfect beams, the nodes are equipped with {\em imperfect} beams that have side-lobe leakage (a scenario that is closer to practice). In this new imperfect 1-2-1 network model, it is still possible to approximate the capacity using collaborative schemes such as Quantize-Map-Forward (QMF)~\cite{QMF,OzgurIT2013} and Noisy Network Coding (NNC)~\cite{NNC} to make use of the multiple access and broadcast  channels present in the network. However, our previous study for the \emph{ideal} 1-2-1 model with perfect main-lobe beamforming beams naturally suggests the following two questions: \emph{(i) When is the ideal 1-2-1 model a good approximation for the imperfect 1-2-1 model? (ii) Under what conditions can simpler schemes involving point-to-point decoding approximate the performance of QMF and NNC in imperfect 1-2-1 networks? }

Our first main result in this paper is to characterize the gap between the approximate capacity of imperfect and  ideal 1-2-1 networks for the same channel coefficients and transmit power. We give sufficient conditions on the parameters of the beamforming pattern for the gap to be constant. Surprisingly, these conditions are independent of the transmit power used by the nodes in the network and only depends on the channel coefficients through a ratio between their values.
Under such conditions,  the ideal 1-2-1 network model offers a good approximation for the imperfect model; we can thus utilize tools developed for the ideal model, such as high-efficiency scheduling algorithms, without incurring significant losses over the imperfect model. 

Our second result explores the gap between the approximate capacity of the ideal 1-2-1 network model and the rate achieved by a simple scheme that consists of decoding point-to-point transmissions while treating side-lobe leakage as noise. We show that we can characterize the aforementioned gap which, different from our first result,
depends on the transmitted power in the network and the individual channel coefficient values.

 
\noindent{\bf Related Work.}
Studies of mmWave communications have focused on profiling the distribution of the Signal-to-Interference-plus-Noise Ratio (SINR) due to side-lobes for single-hop multiple unicast transmissions in cellular and ad-hoc network settings~\cite{thornburg2016performance, bai2014coverage, ding2017random}. Differently, in our setting we are interested in the effect of side-lobe leakage on a single unicast session with multi-hop communication. 
Treating side-lobe reception as noise is closely related to treating interference as noise over the Gaussian interference channel~\cite{motahari2009capacity,geng2015optimality} and in the aforementioned mmWave studies. Differently, in this work we are focused on unicast traffic over a multi-hop network, where all transmissions (including those through side-lobes) are effectively useful to the destination.

\noindent{\bf{Paper Organization.}}
Section~\ref{sec:model} describes the Gaussian ideal and imperfect 1-2-1 network models in full-duplex and presents approximate capacity results.
Section~\ref{sec:imperfect_to_ideal} characterizes the gap between the approximate capacities of the ideal and imperfect 1-2-1 networks.
Finally, Section~\ref{sec:TSN} characterizes the gap between the ideal 1-2-1 model approximate capacity and the rate achieved by treating side-lobe transmissions as noise. Some of the proofs are delegated to the appendices.
~

\section{System Model and Capacity Formulation}
\label{sec:model}
\emph{Notation:} $[n_1:n_2]$ denotes the set of integers from $n_1$ to $n_2 \geq n_1$; $\emptyset$ is the empty set; $\odot$ denotes the Hadamard product and $I$ is the identity matrix. $|\cdot|$ is the absolute value of a scalar as well as the cardinality of a set.
For a matrix $A$: $A^\dagger$ is the complex conjugate of $A$; $[A]_{ij}$ is the element in the $i$-th row and $j$-th column of of $A$; $[A]_j$ is the $j$-th row of $A$.

We consider an $N$-relay Gaussian 1-2-1 network~\cite{EzzeldinISIT2018} where $N$ relays assist the communication between a source node (node~$0$) and a destination node (node~$N+1$).
We  assume full-duplex mode operation for the relays, where each relay $i \in [1:N]$ can be simultaneously receiving and transmitting.
Thus, each node $i \in [0:N+1]$ in the network is characterized by two states, namely $s_{i,t}$ and $s_{i,r}$, that represent the node towards which node $i$ is beamforming its transmissions and the node towards which node $i$ is pointing its receiving beam, respectively.
In particular, $\forall i \in [0:N+1]$, we have that
\begin{align}\label{eq:state}
\begin{array}{ll}
s_{i,t} \subseteq [1:N+1]  \backslash \{i \}, & |s_{i,t}| \leq 1,
\\ s_{i,r} \subseteq [0:N]  \backslash \{i \}, & |s_{i,r}| \leq 1,
\end{array}
\end{align}
where $s_{0,r} = s_{N+1,t} = \emptyset$ since the source node always transmits and the destination node always receives.

{\bf Vanilla 1-2-1 network \cite{EzzeldinISIT2018}:} At  any particular time, a node can only direct (beamform) its transmission towards at most one other node through a {\em perfect} main-lobe beamforming beam with no side-lobes.
Similarly, a node can only receive transmissions from at most another node (to which its receiving main-lobe beam points towards).
Node $j$ receives transmission from node $i$ only if node $i$ points its transmitting beam towards node $j$, and simultaneously, node $j$ points its receiving beam towards node $i$. The channel coefficient between nodes $i$ and $j$ is enhanced by a gain $\alpha > 0$.


{\bf Imperfect 1-2-1 network:} We here introduce the {\em imperfect} 1-2-1 network model, where transmissions/receptions are still achieved by aligning main-lobes, but in addition, transmissions/receptions through side-lobes also occur and are not suppressed as in the ideal model.
In particular, we assume that at any point in time, the channel coefficient $h_{ji}$ from node $i$ to node $j$ is enhanced by a gain $\alpha$ when the main-lobes are aligned, and is attenuated by a factor $\beta$ otherwise. 
Thus, we have the following memoryless channel model 
\begin{subequations}
\label{eq:model_imperfect}
\begin{align}
    &Y_j = Z_j + \sum_{i \in [0:N]\backslash\{j\}} \widehat{h}_{ji}X_i, \quad \forall j \in [1:N+1] \\
    &\widehat{h}_{ji} =
\begin{cases}
\alpha h_{ji} & \text{if $s_{i,t} = \{j\}$, $s_{j,r} = \{i\}$},\\
\beta h_{ji} & \text{otherwise},
\end{cases}
\end{align}
\end{subequations}
where: (i) $s_{i,t}$ and $s_{i,r}$ are defined in~\eqref{eq:state};
(ii) $X_i$ (respectively, $Y_i$) denotes the channel input (respectively, output) at node $i$; 
(iii) $h_{ji} \in \mbb{C}$ is the complex channel coefficient from node $i$ to node $j$ without beamforming; the channel coefficients are assumed to be time-invariant;
(iv) the channel inputs are subject to an individual power constraint, i.e., $\E[|X_k|^2] \leq P,\ k \in [0:N]$;
(v) $Z_j,\ j \in [1:N+1]$ indicates the additive white Gaussian noise at the $j$-th node; noises across the network are assumed to be i.i.d. as $\mcal{CN}(0,1)$.
We use a matrix $H$ to record all the channel coefficients $h_{ji}$ between any two nodes in the network, where the rows are indexed by $[1:N+1]$ and the columns are indexed by $[0:N]$.
\begin{rem}\label{rem:ideal_from_imperfect}
{\rm 
Note that the ideal 1-2-1 network can be recovered from the imperfect model by setting $\beta = 0$. 
}
\end{rem}
The Shannon capacity $\msf{C}$ of the imperfect network in~\eqref{eq:model_imperfect} is not known. However, using similar arguments as in~\cite[Theorem 1]{EzzeldinISIT2018}, it is not difficult to see that $\C$ can be approximated to within a constant gap by the approximate capacity  $\msf{C}_{\rm cs,iid}$ below. In particular, $\msf{C}$ can be upper and lower bounded as
\begin{subequations}
\label{eq:constGap}
\begin{align}
&\msf{C}_{\rm cs,iid} \leq \mathsf{C} \leq \msf{C}_{\rm cs,iid} + O(N\log N),
\\
&\msf{C}_{\rm cs,iid} \!=\!\!\! {\max_{\substack{\lambda_s: \lambda_s \geq 0 \\ \sum_s \lambda_s =1}}} \min_{\substack{\Omega \subseteq[0{:}N],\\ 0 {\in} \Omega}} \sum_s \! \lambda_s {\log}{\det}\!\left(\!I {+} P {H}_{s,\Omega} {H}^\dagger_{s,\Omega}\!\right)\!{,} \label{eq:apprCap}
\\
& {H}_{s,\Omega} = B_s \odot H_\Omega,
\label{eq:GAP}
\end{align}
\end{subequations}
where:
(i) $\Omega$ enumerates all possible cuts in the graph representing the network, such that the source belongs to a set of vertices $\Omega$ and the destination belongs to $\Omega^c$;
(ii) $\Omega^c = [0:N+1] \backslash \Omega$;
(iii) $s$ enumerates all possible network states of the 1-2-1 network in full-duplex, where each network state corresponds to specific values for the variables in~\eqref{eq:state} for each network node; 
(iv) $H_\Omega$ is the submatrix of the channel matrix $H$ selected by retaining only the rows indexed by $\Omega^c$ and the columns indexed by $\Omega$ and reorganizing the rows and columns such that the links that are multiplied by $\alpha$ are along the diagonal; (v) $B_s$ is the beamforming matrix that defines which links are multiplied by the main-lobes gain $\alpha$ and which links are attenuated by $\beta$ in the state $s$; 
(vi) $\lambda_s$, i.e., the optimization variable, is the fraction of time for which state $s$ is active; we refer to a network \emph{schedule} as a collection of $\lambda_s$ for all feasible states, such that they sum up to at most 1.

%
\section{From imperfect to ideal 1-2-1 networks}\label{sec:imperfect_to_ideal}
In this section, we derive an upper bound on the difference between  the approximate capacity of ideal Gaussian 1-2-1 networks (i.e., with $\beta = 0$) which we denote by $\msf{C}_{\rm ideal}$, and the approximate capacity $\msf{C}_{\rm cs,iid}$ of imperfect Gaussian 1-2-1 networks. In particular, $\msf{C}_{\rm ideal}$ is given by~\cite{EzzeldinISIT2018}
\begin{subequations}
\label{eq:constGap_ideal}
\begin{align}
&\msf{C}_{\rm ideal} \!=\! \max_{\substack{\lambda_s: \lambda_s \geq 0 \\ \sum_s \lambda_s = 1}} \min_{\substack{\Omega : \Omega \subseteq [0:N],\\ 0 \in \Omega}} \!\! \sum_{\substack{(i,j): i \in \Omega,\\ j \in \Omega^c}} {\left( \sum_{\substack{ s:\\ j \in s_{i,t},\\ i \in s_{j,r} }} \!\! \lambda_s \right) \ell_{ji}},\label{eq:apprCap_ideal}
\\
& \ell_{ji} = \log\left(1 +P\alpha^2\left| h_{ji}\right|^2\right),
\label{eq:GAP}
\end{align}
\end{subequations}
where $s_{i,t}$ and $s_{i,r}$ denote the transmitting and receiving states for node $i$ in the network state $s$ as defined in~\eqref{eq:state}.

The expression $\msf{C}_{\rm ideal}$ in~\eqref{eq:apprCap_ideal} is appealing as it evaluates the approximate capacity in terms of the point-to-point link capacities $\ell_{ji}$. This, in turn, leads to  interesting properties on how the network should be optimally operated and how to efficiently find an optimal schedule $\{\lambda_s\}$ in polynomial-time in $N$ as discussed in~\cite{EzzeldinISIT2018}. Thus, we would like to understand when the ideal 1-2-1 network model is a good proxy for the imperfect 1-2-1 network. We explore this by characterizing the gap between the approximate capacities of the two  models. 

We characterize the gap between the approximate capacities of the two  models under the following two assumptions that are reasonable for beamforming applications:\\
\noindent $\bullet$ {\bf (Main-lobe is always stronger):}  We assume that $\forall i \neq j \neq k$ such that $h_{ji}, h_{jk} \neq 0$, we have that $\alpha |h_{ji}| \geq \beta |h_{jk}|$.\\
\noindent $\bullet$ {\bf (Each cut is diagonally dominant):} We assume that $\forall s, \Omega$, the matrix $A_{s,\Omega} \!= \! I {+} P {H}_{s,\Omega} {H}^\dagger_{s,\Omega}$ is \emph{diagonally dominant}, i.e.,
\begin{align}\label{eq:rho}
\rho_{s,\Omega}(H) = \max_{i \in [1:|\Omega|^c]} \left\{\sum_{j \in [1:|\Omega|]\backslash\{i\}} \frac{|[A_{s,\Omega}]_{ij}|}{|[A_{s,\Omega}]_{ii}|} \right\} \leq 1.
\end{align}
Our main results are provided by the following two theorems, which are proved in Appendices~\ref{sec:append_gap} and~\ref{sec:append_gap_const}, respectively.
\begin{thm}\label{thm:gap}
Consider an $N$-relay Gaussian 1-2-1 network with channel matrix $H$. Assuming that $(\alpha,\beta)$ are selected such that the two assumptions above are satisfied, then the gap between $\msf{C}_{\rm ideal}$ and $\msf{C}_{\rm cs,iid}$ is upper bounded by
\begin{align}\label{eq:gap_general}
    |\C_{\rm cs,iid} - \C_{\rm ideal}| \leq N \max\left\{\log N,\ f(H,\alpha,\beta)\right\},
\end{align}
where
\begin{align*}
 	f(H,\alpha,\beta) = \max_{s,\Omega} \left| \log\left(1-\rho_{s,\Omega}(H)\right)\right|,
\end{align*} 
with $\rho_{s,\Omega}(H)$ defined in~\eqref{eq:rho}. 
\end{thm}
The gap expressed in Theorem~\ref{thm:gap} depends on the beamforming parameters $(\alpha,\beta)$ and the channel coefficients through the expression $\rho_{s,\Omega}{(H)}$. In order for the ideal 1-2-1 network model to be a valid approximation for the imperfect 1-2-1 network model (i.e., to ensure that the two approximate capacities are a constant gap away), we would like to operate in the range of parameters $(\alpha,\beta)$ such that the gap in Theorem~\ref{thm:gap} is bounded by $N\log(N)$. Our second theorem provides sufficient conditions on $(\alpha,\beta)$, as a function of the channel coefficients, such that $|\C_{\rm cs,iid} - \C_{\rm ideal}| \leq N\log (N)$.
\begin{thm}\label{thm:gap_const}
Consider an $N$-relay Gaussian 1-2-1 network with channel matrix $H$, and let $\Delta$ be the maximum degree of the graph representing the network topology. If the beamforming parameters $(\alpha,\beta)$ satisfy that 
\begin{equation}\label{eq:condition_for_const_gap}
\frac{\alpha}{\beta} \geq \Delta^2 \frac{N}{N-1}\max_{\substack{(i,j,m,n) : |h_{ji}| > 0,\\i\neq j\neq m \neq n}}\frac{|h_{mn}|^2}{|h_{ji}|^2},
\end{equation}
then we have that
$|\C_{\rm cs,iid} - \C_{\rm ideal}| \leq N\log (N)$.
\end{thm}

\begin{rem}
{\rm 
    Note that the condition in~\eqref{eq:condition_for_const_gap} above is independent of the operating power $P$, i.e., it is valid for any operational transmit power used in the network. Furthermore, the condition does not depend on the single channel coefficients, but rather is related to the maximum ratio between the magnitudes of any two non-zero channel coefficients in the network. Thus, for any given network with finite channel coefficient magnitudes, there exists an $(\alpha^\star,\beta^\star)$ pair such that the approximate capacity of the imperfect 1-2-1 network model is at most a constant gap away from the approximate capacity of the ideal 1-2-1 network model that uses $(\alpha^\star,0)$ as described in Remark~\ref{rem:ideal_from_imperfect}.  
}
\end{rem}

\begin{rem}
{\rm 
    The coupling of the approximate capacity of the imperfect 1-2-1 network to the approximate capacity of its ideal 1-2-1 counterpart, allows to translate results already proven for the ideal 1-2-1 network model to the imperfect 1-2-1 network. In particular, in~\cite{EzzeldinISIT2018}, we proved that we can find an optimal schedule for the ideal 1-2-1 network model (i.e., a schedule that achieves $\C_{\rm ideal}$) in polynomial-time in the number of nodes. The proof of Theorem~\ref{thm:gap} and the result in Theorem~\ref{thm:gap_const} imply that, by applying a schedule developed for the ideal model to the imperfect model, we can achieve a rate that is at most a constant gap away from the rate achieved in the ideal model. 
As a result, for any imperfect network where the condition in Theorem~\ref{thm:gap_const} is satisfied, a schedule that is optimal for the ideal model counterpart is also approximately optimal for the imperfect model, i.e., it achieves the capacity with at most a constant gap penalty. Thus, we can leverage the scheduling algorithms in~\cite{EzzeldinISIT2018} to efficiently schedule our imperfect networks.
}
\end{rem}

\section{Treating side-lobe transmissions as noise}
\label{sec:TSN}
The approximate capacity $\msf{C}_{\rm cs,iid}$ for imperfect 1-2-1 networks in~\eqref{eq:constGap} can be achieved using schemes such as QMF~\cite{QMF} and NNC~\cite{NNC}. However, given the relation that we have established between the imperfect and ideal approximate capacities in Theorem~\ref{thm:gap_const}, it is worth exploring how far the rate achieved by simple schemes that rely on point-to-point decoding is from the approximate capacity. 

In this section, we focus on characterizing the gap between the approximate capacity in~\eqref{eq:constGap_ideal} of the ideal 1-2-1 network model and the rate achieved when side-lobe transmissions are treated as noise and only the aligned main-lobes transmissions are decoded at their intended receiver node. 
In particular, for a given network state $s$, node $i$ can communicate to node $j$ at a rate $\widehat{\ell}_{ji}$ given by
\begin{align}\label{eq:TIN_rate_point_point}
\widehat{\ell}_{ji} \!=\! 
\left \{
\begin{array}{ll}
\!\!\!\log\left(1 \!+\! \frac{\alpha^2P|h_{ji}|^2}{1 \!+\! \sum_{m: m\neq i} \beta^2P|h_{jm}|^2}\right) & \text{if }j{\in}s_{i,t}, i{\in}s_{j,r}\!\!\!\!
\\
\!\!\! 0 & \text{otherwise}
\end{array}
\right . .
\end{align}
In other words, the rate $\widehat{\ell}_{ji}$ is either a positive value if the beams are aligned and zero otherwise.
Thus, it is not difficult to see that the maximum rate achieved by this scheme can be computed by considering a ideal 1-2-1 network where the point to point link capacities $\ell_{ji}$ are replaced by $\widehat{\ell}_{ji}$ in~\eqref{eq:TIN_rate_point_point}.
It therefore follows that the achievable rate $\msf{R}_{\rm TSN}$ by Treating Side-lobes as Noise (TSN) is given by
\begin{align}\label{eq:rate_TSN}
\msf{R}_{\rm TSN} = \max_{\substack{\lambda_s: \lambda_s \geq 0 \\ \sum_s \lambda_s = 1}} \min_{\substack{\Omega : \Omega \subseteq [0:N],\\ 0 \in \Omega}} \!\! \sum_{\substack{(i,j): i \in \Omega,\\ j \in \Omega^c}} {\left( \sum_{\substack{ s:\\ j \in s_{i,t},\\ i \in s_{j,r} }} \!\! \lambda_s \right) \widehat{\ell}_{ji}}.
\end{align}
We now focus on characterizing the gap $|\C_{\rm ideal} - \msf{R}_{\rm TSN}|$. If the aforementioned gap is upper bounded by a quantity that only depends on $N$ for some conditions on the beamforming parameters $(\alpha,\beta)$, then by invoking the result in Theorem~\ref{thm:gap_const} together with the triangle inequality, we can show that $|\C_{\rm cs,iid} - \msf{R}_{\rm TSN}|$ is also upper bounded by a constant gap.

It is not difficult to see that $\msf{R}_{\rm TSN} \leq \C_{\rm ideal}$, since for all $i,j$, we have $\widehat{\ell}_{ji} \leq {\ell}_{ji}$. For a lower bound on $\msf{R}_{\rm TSN}$, we can use the following lower bound on $\widehat{\ell}_{ji}$
\begin{align}
\widehat{\ell}_{ji} &= \log\left(1 + \alpha^2P|h_{ji}|^2 + \sum_{m: m\neq i} \beta^2P|h_{jm}|^2\right) \nonumber \\
&\quad - \log\left(1 + \sum_{m: m\neq i} \beta^2P|h_{jm}|^2\right) \nonumber \\
&\geq  \log\left(1 + \alpha^2P|h_{ji}|^2\right)\nonumber\\ 
&\quad - \underbrace{\log\left(1 + \max_{m: m\neq i} \beta^2P|h_{jm}|^2\right)}_{\widetilde{\ell}_{ji}} - \log(\Delta) \nonumber \\
&\geq \ell_{ji} - \max_{i,j}\widetilde{\ell}_{ji} - \log(\Delta),
\end{align} 
where $\Delta$ is the maximum degree of the graph representing the network topology.
We can then lower bound  $\msf{R}_{\rm TSN}$ as follows
\begin{align}
\msf{R}_{\rm TSN} &= \max_{\substack{\lambda_s: \lambda_s \geq 0 \\ \sum_s \lambda_s = 1}} \min_{\substack{\Omega : \Omega \subseteq [0:N],\\ 0 \in \Omega}} \!\! \sum_{\substack{(i,j): i \in \Omega,\\ j \in \Omega^c}} {\left( \sum_{\substack{ s:\\ j \in s_{i,t},\\ i \in s_{j,r} }} \!\! \lambda_s \right) \widehat{\ell}_{ji}} \nonumber \\
&\geq \max_{\substack{\lambda_s: \lambda_s \geq 0 \\ \sum_s \lambda_s = 1}} \min_{\substack{\Omega : \Omega \subseteq [0:N],\\ 0 \in \Omega}} \!\! \sum_{\substack{(i,j): i \in \Omega,\\ j \in \Omega^c}} {\left( \sum_{\substack{ s:\\ j \in s_{i,t},\\ i \in s_{j,r} }} \!\! \lambda_s \right) {\ell}_{ji}}\nonumber \\
&\quad - N\log(\Delta) - N\max_{i,j} \widetilde{\ell}_{ji}.
\end{align}
As a result, we have
\begin{align}\label{eq:final_gap_TSN}
|\C_{\rm ideal} - \msf{R}_{\rm TSN}| \leq N\log(\Delta) + N\max_{i,j} \widetilde{\ell}_{ji}.
\end{align}
Note that, for $\msf{R}_{\rm TSN}$ to be a constant gap away from $\msf{C}_{\rm ideal}$, $\beta$ should be selected such that for all channel coefficients in the network we have that $\widetilde{\ell}_{ji} = O(1)$.

\begin{rem}
{\rm 
The conditions imposed by~\eqref{eq:final_gap_TSN} and Theorem~\ref{thm:gap_const} on the beamforming parameters are realistic for several envisioned applications of mmWave communications. For example, in a typical vehicle platooning scenario~\cite{zeng2019joint}, the inter-platoon distance is around $10$ meters. Thus, with an operating frequency of $60$ GHz and bandwidth of $1$ GHz, the largest channel coefficient magnitude $\max_{ij} |h_{ij}| = O(10^{-4})$. Assuming the line-of-sight path loss model and a transmit signal-to-noise ratio of $100$, we would have $P|h_{ij}|^2 = O(10^{-2})$. As a result, even with $\beta = 1$, the gap in~\eqref{eq:final_gap_TSN} would be upper bounded by $N[\log(\Delta) + 1]$.
}
\end{rem}

\appendices
\section{Proof of Theorem~\ref{thm:gap}}\label{sec:append_gap}
To prove Theorem~\ref{thm:gap}, we focus on $\C(s,\Omega)$ below for all valid states $s$ and cuts $\Omega$
\begin{equation}\label{eq:cut_state_value}
\C(s,\Omega) = \log\det\left(I + P{H}_{s,\Omega} {H}^\dagger_{s,\Omega}\right).
\end{equation}
We seek to understand how this term relates to its counterpart in the expression for $\C_{\rm ideal}$ in~\eqref{eq:constGap_ideal}.
Without loss of generality, we assume that for the given $s$, $\Omega$, the considered ${H}_{s,\Omega}$ matrix is wide, otherwise, we can consider the expression in~\eqref{eq:cut_state_value} with the conjugate transpose matrix.

We can get an upper bound on $\C(s,\Omega)$ in~\eqref{eq:cut_state_value} as follows
\begin{align}
\label{eq:upperbound_cut}
\C(s,\Omega) &= \log\det\left(I + P{H}_{s,\Omega} {H}^\dagger_{s,\Omega}\right) \nonumber \\
             &\stackrel{{\rm{(a)}}}\leq \log\left(\prod_{j=1}^{|\Omega^c|} \left[I + P{H}_{s,\Omega} {H}^\dagger_{s,\Omega}\right]_{jj}\right) \nonumber\\
             &= \sum_{j=1}^{|\Omega^c|} \log\left( 1 + P\left\|\left[{H}_{s,\Omega}\right]_{j}\right\|^2\right) \nonumber
             \\
             &\leq \sum_{j=1}^{|\Omega^c|} \log\left( 1 + P\max_{i \in [1:|\Omega|]} \left| \left[{H}_{s,\Omega}\right]_{ji}\right|^2\right) \nonumber
             \\
             &\quad + |\Omega^c| \log(|\Omega|),
\end{align}
where $\rm{(a)}$ follows from the Hadamard-Fischer inequality~\cite{rozanski2017more} that upper bounds the determinant of a positive semidefinite matrix with the product of its diagonal elements.

By using our ``main-lobe is always stronger'' assumption, i.e.,
$\alpha$ and $\beta$ are such that the side-lobe transmissions are weaker than those on the main-lobe, we can simplify~\eqref{eq:upperbound_cut} as
\begin{align}\label{eq:upperbound_cut_state}
\C(s,\Omega) &\leq \sum_{j=1}^{|\Omega^c|} \log\left( 1 {+} P\left|\left[{H}_{s,\Omega}\right]_{jj}\right|^2\right) \nonumber
\\
&\quad + |\Omega^c| \log(|\Omega|).
\end{align}
Recall that each entry $\left[{H}_{s,\Omega}\right]_{jj}$ in the matrix ${H}_{s,\Omega}$ corresponds to the enhanced channel coefficient between two nodes that have their main-lobe beams aligned.
Thus, we have the following upper bound on the approximate capacity $\C_{\rm cs,iid}$ in~\eqref{eq:apprCap}
\begin{align}
&\C_{\rm cs,iid} = \max_{\substack{\lambda_s: \lambda_s \geq 0 \\ \sum_s \lambda_s = 1}} \min_{\substack{\Omega \subseteq[0:N],\\ 0 \in \Omega}} \C(s,\Omega) \nonumber \\
                &\leq N\log(N) {+} \max_{\substack{\lambda_s: \lambda_s \geq 0 \\ \sum_s \lambda_s = 1}} \min_{\substack{\Omega \subseteq[0:N],\\ 0 \in \Omega}} 
                \sum_s \lambda_s \sum_{\substack{(i,j) \in \Omega \times \Omega^c:\\ j \in s_{i,t},\ i \in s_{j,r}}} \ell_{ji} \nonumber \\
                &= N\log(N) + \C_{\rm ideal}.
\end{align}
We now want to find a lower bound for $\C_{\rm cs,iid}$ in terms of $\C_{\rm ideal}$. To do this, we will again focus on each of the terms $\C(s,\Omega)$ in~\eqref{eq:cut_state_value}.
Recall that,  by our ``each cut is diagonally dominant'' assumption, we have that for all $s$, $\Omega$, the matrix $I + P H_{s,\Omega}H^\dagger_{s,\Omega}$ is diagonally dominant.

For a diagonally dominant $n\times n$ matrix $A$, we can use the result in~\cite{ostrowski1952note} to derive the following lower bound on the determinant of $A$
\begin{align}\label{eq:lowerbound_det}
\det(A) &\stackrel{\text{\cite{ostrowski1952note}}}{\geq} \prod_{i=1}^n \left( [A]_{ii} - \sum_{j \in [1:n]\backslash\{i\}} |[A]_{ij}|\right) \nonumber \\
       &\geq \prod_{i=1}^n \left([A]_{ii} - \rho_A [A]_{ii}\right) = (1-\rho_A)^n \prod_{i=1}^n A_{ii},
\end{align}
where $\rho_A$ is given by 
\begin{align}
\rho_A = \max_{i \in [1:n]} \left\{\sum_{j \in [1:n]\backslash\{i\}} \frac{|[A]_{ij}|}{|[A]_{ii}|} \right\}.
\end{align}
Now, by employing~\eqref{eq:lowerbound_det} on the matrix $I + P{H}_{s,\Omega} {H}^\dagger_{s,\Omega}$ in~\eqref{eq:cut_state_value}, we have that
\begin{align}
&\C(s,\Omega) = \log\det\left(I + P{H}_{s,\Omega} {H}^\dagger_{s,\Omega}\right) \nonumber \\
             &\geq \sum_{j=1}^{|\Omega^c|} \log\left( 1 + P\left\|\left[{H}_{s,\Omega}\right]_j\right\|^2\right) + \log\left((1-\rho_{s,\Omega}(H))^{|\Omega^c|}\right) \nonumber \\
             &\geq \sum_{j=1}^{|\Omega^c|} \log\left( 1 + P \left| \left[{H}_{s,\Omega}\right]_{jj}\right|^2\right)
              - |\Omega^c|\left|\log(1-\rho_{s,\Omega}(H))\right|\nonumber \\
             &\geq \sum_{j=1}^{|\Omega^c|} \log\left( 1 {+} P \left| \left[{H}_{s,\Omega}\right]_{jj}\right|^2\right)
              {-} N {\underbrace{\left|\log(1\!-\!\rho_{s,\Omega}(H))\right|}_{f(\rho,s,\Omega)}}.
\end{align}
Thus, we have a lower bound on the approximate capacity as follows
\begin{align}\label{eq:lowerbound_cut}
&\C_{\rm cs,iid} = \max_{\substack{\lambda_s: \lambda_s \geq 0 \\ \sum_s \lambda_s = 1}} \min_{\substack{\Omega \subseteq[0:N],\\ 0 \in \Omega}} \C(s,\Omega) \nonumber \\
                &\geq \max_{\substack{\lambda_s: \lambda_s \geq 0 \\ \sum_s \lambda_s = 1}} \min_{\substack{\Omega \subseteq[0:N],\\ 0 \in \Omega}} 
                \sum_s \lambda_s \left(\sum_{\substack{(i,j) \in \Omega \times \Omega^c:\\ j \in s_{i,t},\ i \in s_{j,r}}} \ell_{ji} - N f(\rho,s,\Omega)\right) \nonumber \\ 
                &\geq \C_{\rm ideal} - N\max_{s,\Omega}f(\rho,s,\Omega) \nonumber \\
                &= \C_{\rm ideal} - N\left|\log(1-\rho_{s,\Omega}(H))\right|.
\end{align}
By taking the maximum among the gaps in the bounds in~\eqref{eq:upperbound_cut} and~\eqref{eq:lowerbound_cut}, we get the result in Theorem~\ref{thm:gap}.
\section{Proof of Theorem~\ref{thm:gap_const}}\label{sec:append_gap_const}
To prove Theorem~\ref{thm:gap_const}, we would like to derive bounds on the pair $(\alpha,\beta)$ such that $f(H,\alpha,\beta)$ in~\eqref{eq:gap_general} is upper bounded by $\log(N)$. By simple arithmetic manipulation, it is not difficult to see the following equivalence 
\begin{align}
\label{eq:SuffCond}
f(H,\alpha,\beta) &= \max_{s,\Omega} \left| \log\left(1-\rho_{s,\Omega}(H)\right)\right| \leq \log(N) \nonumber \\ 
&\iff \max_{s,\Omega} \rho_{s,\Omega}(H) \leq \frac{N-1}{N}.
\end{align}
We now find an upper bound on $\rho_{s,\Omega}(H)$ and then derive the sufficient condition in Theorem~\ref{thm:gap_const} by enforcing that the upper bound on $\rho_{s,\Omega}(H)$ is less than or equal to $(N-1)/N$, $\forall s,\Omega$.

From the definition of $\rho_{s,\Omega}(H)$ in~\eqref{eq:rho}, we have that
\begin{align}\label{eq:upperbound_1}
\rho_{s,\Omega}(H)
&= \max_{i \in [1:|\Omega^c|]}\left\{ \sum_{\substack{j \in [1:|\Omega|]\setminus\{i\}}} \frac{\left|P\left[{H}_{s,\Omega}\right]_i\left[{H}_{s,\Omega}\right]^\dagger_j\right|}{1+P\left[{H}_{s,\Omega}\right]_i\left[{H}_{s,\Omega}\right]_i^\dagger}\right\} \nonumber \\
&\leq \max_{i \in [1:|\Omega^c|]}\left\{ \sum_{\substack{j \in [1:|\Omega|]\setminus\{i\}}} \frac{\left|P\left[{H}_{s,\Omega}\right]_i\left[{H}_{s,\Omega}\right]^\dagger_j\right|}{P\alpha^2 \left|\left[{H}_{s,\Omega}\right]_{ii}\right|^2}\right\} \nonumber \\
&= \max_{i \in [1:|\Omega^c|]}\left\{ \sum_{\substack{j \in [1:|\Omega|]\setminus\{i\}}} \frac{\left|\left[{H}_{s,\Omega}\right]_i\left[{H}_{s,\Omega}\right]^\dagger_j\right|}{\alpha^2\left|\left[{H}_{s,\Omega}\right]_{ii}\right|^2}\right\}.
\end{align}
Now, note that
\begin{align}\label{eq:dot_prod_upperbound}
&\left|\left[{H}_{s,\Omega}\right]_i\left[{H}_{s,\Omega}\right]^\dagger_j\right|
{\stackrel{\rm (a)}=} \left| \alpha\beta(\widehat h_{ii}\widehat h^*_{ji} {+} \widehat h_{ij}\widehat h^*_{jj}) {+} \beta^2 \sum_{k \neq i,j} \widehat h_{ik}\widehat h^*_{jk} \right| \nonumber \\
&\stackrel{\rm (b)}\leq \left[2\alpha\beta + (\Delta-2)\beta^2\right]\max_k\left\{\max\{ |\widehat h_{ik}|^2, |\widehat h_{jk}|^2\}\right\} \nonumber \\
&\leq \left[2\alpha\beta + (\Delta-2)\beta^2\right]\max_{i,j} |h_{ij}|^2,
\end{align}
where: $\rm{(a)}$ uses $\widehat h_{ij} = [H_{s,\Omega}]_{ij}$ as in~\eqref{eq:model_imperfect} for space limitation; $\rm{(b)}$ follows from the triangle inequality and the fact that, in the dot product $\left[{H}_{s,\Omega}\right]_i\left[{H}_{s,\Omega}\right]_j^\dagger$, there are at most $\Delta$ non-zero terms since $\Delta$ is the maximum degree of the graph representing the network topology. 

By substituting~\eqref{eq:dot_prod_upperbound} in~\eqref{eq:upperbound_1}, we have that 
\begin{align}\label{eq:upperbound_2}
&\rho_{s,\Omega}(H)\nonumber {\leq} \max_{i \in [1:|\Omega^c|]}\left\{ \sum_{\substack{j \in [1:|\Omega|]\setminus\{i\}}} \frac{\left|\left[{H}_{s,\Omega}\right]_i\left[{H}_{s,\Omega}\right]^\dagger_j\right|}{\alpha^2\left|\left[{H}_{s,\Omega}\right]_{ii}\right|^2}\right\} \nonumber \\
&{\leq} {\max_{i \in [1:|\Omega^c|]}}\left\{ {\sum_{\substack{j \in [1:|\Omega|]\setminus\{i\}}}} \frac{\left[2\alpha\beta + (\Delta-2)\beta^2\right]\displaystyle\max_{m,n} |h_{mn}|^2}{\alpha^2\left|\left[{H}_{s,\Omega}\right]_{ii}\right|^2}\right\} \nonumber \\
&{\leq} \frac{\left[2\alpha\beta {+} (\Delta{-}2)\beta^2\right](\Delta{-}1)}{\alpha^2}\max_{\substack{(i,j,m,n) : |h_{ji}| > 0,\\i\neq j\neq m \neq n}} \frac{|h_{mn}|^2}{|h_{ij}|^2}.
\end{align}
Given the upper bound on $\rho_{s,\Omega}(H)$ in~\eqref{eq:upperbound_2} that is independent of $\Omega,\ s$, we can now~\eqref{eq:SuffCond} and get the sufficient condition
\begin{align*}
&\frac{\left[2\alpha\beta {+} (\Delta{-}2)\beta^2\right](\Delta{-}1)}{\alpha^2}\max_{\substack{(i,j,m,n) : |h_{ji}| > 0,\\i\neq j\neq m \neq n}} \frac{|h_{mn}|^2}{|h_{ij}|^2} \leq \frac{N-1}{N}
\\ 
& \Longrightarrow\frac{\alpha}{\beta} \geq \Delta^2\frac{N}{N-1} \max_{\substack{(i,j,m,n) : |h_{ji}| > 0,\\i\neq j\neq m \neq n}} \frac{|h_{mn}|^2}{|h_{ij}|^2}.
\end{align*}
This concludes the proof of Theorem~\ref{thm:gap_const}.
\bibliographystyle{IEEEtran}
\bibliography{mmWaveNetwork}
\end{document}